\documentclass[preprint,showpacs,aps]{revtex4}
\usepackage{graphicx}

\begin{document}

\title{Branching ratios of $\alpha$-decay to excited states of even-even nuclei}
\author{Chang Xu$^{1}$ and Zhongzhou Ren$^{1,2,3} \email{ zren@nju.edu.cn}$ }
\address{$^{1}$Department of Physics, Nanjing University,
Nanjing 210008, China\\
$^{2}$Center of Theoretical Nuclear Physics, National Laboratory
of Heavy-Ion Accelerator, Lanzhou 730000, China \\
$^{3}$CPNPC,  Nanjing University, Nanjing 210008, China}

\begin{abstract}
We make a systematic calculation on $\alpha$-decay branching
ratios to members of ground-state rotational band and to excited
$0^{+}$ states of even-even nuclei by a simple barrier penetration
approach. The branching ratios to the excited states of daughter
nucleus are determined by the $\alpha$-decay energy, the angular
momentum of $\alpha$-particle, and the excitation probability of
the daughter nucleus. Our calculation covers isotopic chains from
Hg to Fm in the mass regions 180$<$A$<$202 and A$\geq$224. The
calculated branching ratios of $\alpha$-transitions are consistent
with the experimental data. Some useful predictions are made for
future experiments.
\end{abstract}


\maketitle

\section{Introduction}

The first successful application of quantum mechanics to nuclear
physics problem is the theory of $\alpha$-decay which was
presented independently by Gamow and by Condon and Gurney in 1928
\cite{gam,con}. Based on the Gamow theory, the experimental
$\alpha$-decay half-lives of nuclei can be well explained by both
phenomenological and microscopic models
\cite{vio,poe,hat,bro,buc,roy,den,moh,xu1,xu2}. Such
$\alpha$-decay calculations are mainly concentrated on the favored
cases, \textit{e.g.} the ground-state to ground-state
$\alpha$-transitions of even-even nuclei ($\triangle$$\ell$=0)
\cite{fir,mol}. Besides the favored $\alpha$-transitions, the
ground-state of the parent nucleus can also decay to the excited
states of the daughter nucleus ($\triangle$$\ell$$\neq$0)
\cite{fir}. Recently, there is increasing interest in two kinds of
$\alpha$-transitions of even-even nuclei from both experimental
and theoretical sides, \textit{i.e.} the $\alpha$-decay to excited
0$^{+}$ states and to members of the ground-state rotational band
\cite{wau,and,ric,del,kar,asa}. These $\alpha$-transitions belong
to the unfavored case, which are strongly hindered as compared
with the ground-state ones. Theoretically, the hindered
$\alpha$-transition is an effective tool to study the properties
of $\alpha$-emitters because it is closely related to the internal
structure of nuclei \cite{wau,and,ric,del,kar,asa}. However, it is
very difficult to describe quantitatively the unfavored
$\alpha$-transitions due to the influence of both non-zero angular
momentum and excitation of nucleons, especially for
$\alpha$-emitters in the neighborhood of the shell closures.
Although the favored $\alpha$-decay model can be straightforwardly
applied to the unfavored $\alpha$-transition, the calculated
branching ratios usually deviate significantly from the
experimental data. Considering the complexity of hindered
$\alpha$-decay, it is very interesting to find a simple way of
explaining the available data of hindered $\alpha$-decay based on
the favored $\alpha$-decay theory. Experimentally it is also very
helpful to make theoretical predictions on unobserved hindered
$\alpha$-transitions for future studies.

The aim of this paper is to study the hindered
$\alpha$-transitions of even-even nuclei with mass numbers
180$<$A$<$202 and A$\geq$224. We apply a simple barrier
penetration approach to calculate the branching ratios of $\alpha$-decay
based on the Gamow theory. The influence of the $\alpha$-decay
energy, the angular momentum of the $\alpha$-particle, and the
excitation probability of the daughter nucleus are properly taken
into account. The outline of this paper is as follows. Section II
is the framework of the barrier
penetration approach.
The numerical results and
corresponding discussions are given in Section III.
Section IV is a brief summary.

\section{Formalism}

Firstly, we start with the radial Schr\"{o}dinger equation
\begin{eqnarray}
 - \frac{{\hbar ^2 }}{{2\mu }}\frac{{d^2 \psi (r)}}{{dr^2 }} +
 [U(r) + \frac{{\hbar ^2 }}{{2\mu }}\frac{{\ell(\ell + 1)}}{{r^2 }}]\psi (r) = E\psi(r),
\end{eqnarray}
where the centrifugal potential $\frac{{\hbar ^2 }}{{2\mu
}}\frac{{\ell(\ell + 1)}}{{r^2 }}$ is included in the
Schr\"{o}dinger equation and $U(r)$ is the standard square well
potential
\begin{eqnarray}
U(r) = \left\{
       \begin{array}{ll}
       - U_0                 &  (r < R_0 )\\
       Z_1 Z_2 e^2 / r &  (r \ge R_0 ).
 \end{array} \right.
\end{eqnarray}
\begin{figure}[htb]
\centering
\includegraphics[width=8cm]{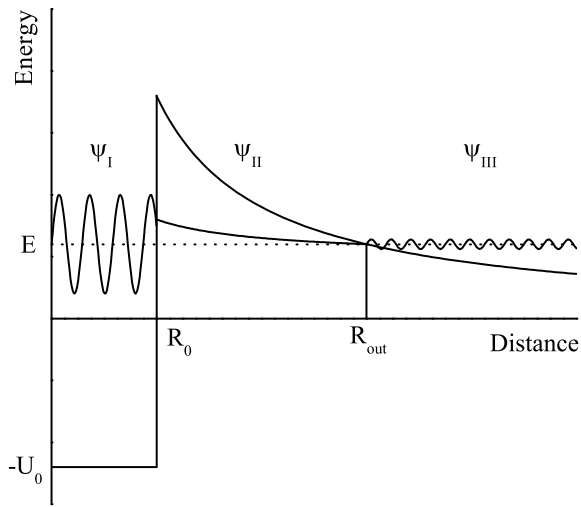}
\caption{The standard square well potential of $\alpha$-decay
and the quantum-tunnelling effect of the $\alpha$-particle.}
\end{figure}

The schematic representation of the square well potential is given
in Fig.1. The $\alpha$-particle in the parent nucleus is initially
trapped within this square well potential and the corresponding
wave function is denoted by $\psi_{I}$ in Fig.1, \textit{i.e.} the
incoming wave function. Through quantum-tunnelling effect the
$\alpha$-particle is finally emitted from the parent nucleus and
then characterized by the wave function of a free particle
$\psi_{III}$, \textit{i.e.} the outgoing wave function. The
penetration probability of the $\alpha$-particle through the
Coulomb barrier is proportional to the square of the ratio between
the outgoing and incoming wave functions. Using the well known WKB
technique, one can obtain the penetration probability of the
$\alpha$-particle \cite{xu}
\begin{eqnarray}
P_\alpha (Q_{\alpha}, E^*_{\ell}, \ell ) \propto
|\frac{\psi_{III}}{\psi_{I}}|^2 = \textrm{exp}[ - 2\int_{R_0
}^{R_{out}}k(r)dr],
\end{eqnarray}
with
\begin{eqnarray}
k(r)={\sqrt{\frac{{2\mu }}{{\hbar ^2 }}}}\,[\frac{{Z_1
Z_2 e^2 }}{r} + \frac{{\hbar ^2 }}{{2\mu }}\frac{{\ell (\ell  +
1)}}{{r^2 }} - (Q_{\alpha}-E^*_{\ell})]^{\frac{1}{2}},
\end{eqnarray}
where $Z_{1}$ and $Z_{2}$ are the charge numbers of the
$\alpha$-particle and the daughter nucleus, respectively. $\mu$ is
the reduced mass of the $\alpha$-core system and $\ell$ is the
angular momentum carried by the $\alpha$-particle. $Q_{\alpha}$ is
the decay energy of the ground-state transition and $E^*_{\ell}$
is the excitation energy of state $\ell$. $R_{0}$ is the radius of
the daughter nucleus ($R_{0}=1.2A_{2}^{1/3}$) and $R_{out}$ is the
outer classic turning point (see Fig.1).

In describing $\alpha$-decay of heavy nuclei, the height of the
centrifugal barrier at $r=R_0$ is generally very small compared
with the Coulomb barrier \cite{gam2}
\begin{eqnarray}
\varepsilon=\frac{{\hbar ^2 }}{{2\mu }}\frac{{\ell (\ell  +
1)}}{{R_0^2 }}:\frac{{Z_1 Z_2 e^2 }}{R_0}.
\end{eqnarray}

By expanding the wave number $k(r)$ in powers of the small
quantity $\varepsilon$, the penetration probability can be written
in a simple form \cite{gam2}
\begin{eqnarray}
P_\alpha (Q_{\alpha}, E^*_{\ell}, \ell ) = \textrm{exp}[
-\sqrt{\frac{2\mu}{\hbar^2}} \frac{Z_1 Z_2 e^{2}
\pi}{(Q_{\alpha}-E^*_{\ell})^{\frac{1}{2}}}\,] \times
\textrm{exp}[ -\sqrt{\frac{ \hbar^2 }{ 2\mu }} \frac{2 \ell(\ell +
1)}{(Z_1 Z_2 e^{2} R_0)^{\frac{1}{2}}}\,],
\end{eqnarray}
where the first term represents the influence of the excitation
energy $E^*_{\ell}$ on the penetration factor and the second term
denotes the influence of the non-zero angular momentum $\ell$.
Historically, the separable form of the penetration factor
(Eq.(6)) was first derived by Gamow \cite{gam2} and then discussed
by Rasmussen \textit{et al.} \cite{ras}. Similar expression of the
penetration factor is also given in the textbook of Bohr and
Mottelson \cite{boh}. From Eq.(6), it is easy to find that the
penetration probability of the $\alpha$-particle reaches maximum
value when both the excitation energy $E^*_{\ell}$ and the angular
momentum $\ell$ are zero, \textit{i.e.} the $\alpha$-transition to
the ground-state of the daughter nucleus. This is in accord with
the experimental fact that the ground-state branching ratio of
$\alpha$-decay is the largest for all even-even $\alpha$-emitters
\cite{fir}. The residual daughter nucleus after disintegration has
the most probability to stay in its ground state, and the
probability to stay in its excited state is relatively much
smaller. Therefore it is a reasonable assumption that the
probability of the residual daughter nucleus to stay in its
excited states ($I^{+}$=2$^{+}$, 4$^{+}$, 6$^{+}$,...) obeys the
Boltzmann distribution
\begin{eqnarray}
w_{{\ell}}(E^*_{\ell})=\textrm{exp}[-cE^*_{\ell}],
\end{eqnarray}
where $E^*_{{\ell}}$ is the excitation energy of state $\ell$ and
$c$ is a free parameter.
We include the excitation probability function (Eq.(7))
in our approach. The value of parameter $c$ is fixed to
1.5 in present calculation. This means that only a single
parameter is introduced in the whole calculation.
It is stressed that the inclusion
of the excitation probability is reasonable in physics and it can lead to
good agreement between experiment and theory.
Here we define $I_{\ell +}$ as the product of the penetration
factor and the excitation probability
\begin{eqnarray}
I_{\ell ^+} = w_{\ell}(E^*_{\ell}) P_\alpha (Q_{\alpha},
E^*_{\ell}, \ell ),
\end{eqnarray}
which denotes the total probability of $\alpha$-transition from
the ground-state of the parent nucleus to the excited state $\ell
^+$ of the daughter nucleus. It is written in a very accessible
style of three exponential factors which contain the essential
theory of $\alpha$-decay. It is very convenient to estimate the
influence of these factors on the hindered $\alpha$-transitions
from $I_{\ell ^+}$. With the help of $I_{\ell ^+}$, the branching
ratios of $\alpha$-decay to each state of the rotational band of
the daughter nucleus can be written as
\begin{eqnarray}
&&b_{g.s.}^{0 ^{+}}\%=I_{0^+}/(I_{0^+}+I_{2^+}+
I_{4^+}+I_{6^+}+...)\times 100\%
\\ \nonumber
&&b_{e.s.}^{2 ^{+}}\%=I_{2^+}/(I_{0^+}+I_{2^+}+
I_{4^+}+I_{6^+}+...)\times 100\%
\\ \nonumber
&&b_{e.s.}^{4 ^{+}}\%=I_{4^+}/(I_{0^+}+I_{2^+}+
I_{4^+}+I_{6^+}+...)\times 100\%
\\ \nonumber
&&....
\end{eqnarray}

Similarly, the branching ratio of $\alpha$-decay to the excited
$0^{+}$ state of the daughter nucleus is given by
\begin{eqnarray}
b_{e.s.}^{0 ^{+}}\%=b_{g.s.}^{0 ^{+}}\% \times
\frac{w_{0}(E^*_{0})}{w_{0}(0)} \frac{P_\alpha (Q_{\alpha}, E^*_{0}, 0 )}
{P_\alpha (Q_{\alpha}, 0, 0)},
\end{eqnarray}
where $b_{g.s.}^{0 ^{+}}$\% is the branching ratio of
$\alpha$-transition between the ground states. It can be further
simplified because the angular momentum carried by the
$\alpha$-particle is zero
\begin{eqnarray}
b_{e.s.}^{0 ^{+}}\%=b_{g.s.}^{0 ^{+}}\% \times
\textrm{exp}[-cE^*_{\ell}] \times \textrm{exp}\{
\sqrt{\frac{2\mu}{\hbar^2}}\, Z_1 Z_2 e^{2}\,
\pi[\frac{1}{Q_{\alpha}^{\frac{1}{2}}}-\frac{1}{(Q_{\alpha}-E^*_{\ell})^{\frac{1}{2}}}]\}.
\end{eqnarray}
The $\alpha$-transition to the excited $0^+$ state of the daughter
nucleus does not involve the change of angular momentum $\ell$,
which is an ideal case for theoretical studies of hindered
$\alpha$-transitions.

Here we would like to mention that the derivation of the above
formulas starts with a square well potential, which is a rough
approximation of the real potential between the $\alpha$-particle
and the daughter nucleus. In principle, the nuclear and Coulomb
potentials determining the $\alpha$-decay process should be very
smooth and not peaked. Actually more realistic potentials have
been widely used to describe the decay properties of a large
number of nuclei. For instance, Royer has made a systematic
calculation on the $\alpha$-decay half-lives by the generalized
liquid drop model (GLDM) \cite{roy}. The important proximity
energy is properly included due to the attractive forces in the
neck and the gap between the two close fragments \cite{roy}. The
obtained $\alpha$-decay barrier is more realistic as compared with
the pure Coulomb barrier \cite{roy}. We also systematically
calculated the favored $\alpha$-decay half-lives with a
double-folding potential in the framework of the density-dependent
cluster model (DDCM) \cite{xu1,xu2}. The nuclear potential from
the double-folding formulism is also microscopic because it
correctly includes the low-density behavior of the nucleon-nucleon
interaction and guarantees the antisymmetrization of identical
particles in the $\alpha$-cluster and in the core \cite{xu1,xu2}.
In present study, we are interested in calculating the
$\alpha$-decay branching ratio to the rotational members and to
the excited 0$^+$ states. To simplify the above problem, we assume
that the nuclear potential vanishes outside the radius of the
daughter nucleus $R_0$ (see Fig.1) by using a square well
potential. However, it is expected that the final expression is
not very sensitive to the form of the nuclear potential. This is
due to the magnitude of the branching ratio is determined by the
proportion of the total penetration factors between the excited
and ground states ($b_{e.s.}^{\ell ^{+}}$= $b_{g.s.}^{0
^{+}}$$I_{\ell ^+}/ I_{0^+}$). In this situation, the influence of
the non-vanishing nuclear potential on the branching ratio can be
approximately cancelled in calculations. This is also different
from the studies of $\alpha$-decay half-lives where the details of
the nuclear potential become very important (decay width $\lambda
\propto I_{\ell ^+}$). It is considered that the present
approximation with a square well potential will not affect the
final results of the branching ratios significantly.

\section{Numerical results and discussions}

The $\alpha$-transitions to members of the ground-state rotational
band are systematically calculated for well deformed nuclei with
Z$\geq$90. The systematic calculation on $\alpha$-decay branching
ratios to the rotational band is rare because some data of the
excited states have been obtained very recently \cite{fir}.
Experimentally it is known that the ground-state of the even-even
actinides mainly decays to the 0$^+$ and 2$^+$ states of their
daughter nucleus \cite{fir}. The sum of branching ratios to the
0$^+$ and 2$^+$ states is as large as 99\% in many cases. The
$\alpha$-transitions to other members of the rotational band
(I$^+$=4$^+$, 6$^+$, 8$^+$...) are strongly hindered. This is
different from the $\alpha$-transition to the ground or excited
$0^+$ state where the angular momentum carried by the
$\alpha$-particle is zero ($\triangle\ell$=0). Here the influence
of the non-zero angular momentum should be included for
$\triangle\ell$$\neq$0 transitions. In Fig.2$-$Fig.4, we give
three typical figures for $\alpha$-decay fine structure of
$^{238}$Pu, $^{242}$Cm and $^{246}$Cf. The $\alpha$-decay
branching ratios of $^{238}$Pu and $^{242}$Cm have been measured
up to 8$^+$ state of the rotational band, and the branching ratio
of $^{246}$Cf has been measured up to 6$^+$ state in experiment.
It is shown in these figures that the calculated values agree with
the experimental ones for both the low-lying states (0$^+$, 2$^+$)
and the high-lying ones (6$^+$, 8$^+$), however, the calculated
branching ratio to 4$^+$ state is slightly larger than the
experimental one. Let us take the $\alpha$-decay of $^{238}$Pu as
an example to illustrate the discrepancy of branching ratio to
$4^+$ state. From Fig.2, we can see that the proportion between
the experimental branching ratios of 2$^+$ and 4$^+$ states is
$b_{e.s.}^{4 ^{+}}\%/b_{e.s.}^{2
^{+}}\%$=28.98\%/0.105\%$\approx$276, but the proportion between
the branching ratios of 4$^+$ and 6$^+$ states is only
0.105\%/0.0030\%$\approx$35 and the proportion between 6$^+$ and
8$^+$ states is also small
0.0030\%/(6.8$\times$10$^{-5}$\%)$\approx$44. Thus the proportion
of $b_{e.s.}^{4 ^{+}}\%/b_{e.s.}^{2 ^{+}}\%$ is very large as
compared with those of other rotational members. This shows that
the variation of the experimental branching ratios is not very
smooth. On the contrary, our calculated branching ratios vary
smoothly for different excited states and this leads to the slight
disagreement between theory and data for $4^+$ state. Besides the
calculation of decay chain
$^{246}$Cf$\rightarrow$$^{242}$Cm$\rightarrow$$^{238}$Pu, we have
also calculated the branching ratios of ground-state rotational
band for even-mass $\alpha$-emitters $^{224-230}$Th,
$^{228-238}$U, $^{236-244}$Pu, $^{240-248}$Cm, $^{240-252}$Cf and
$^{250-254}$Fm. The discrepancy in describing 4$^+$ state also
exists for these even-even $\alpha$-emitters in our calculation.
We call this as the abnormity of 4$^+$ state in $\alpha$-decay for
convenience. It is very interesting to pursue this by performing
more microscopic calculation in future. Nevertheless, the overall
agreement of branching ratios to the rotational band of these
nuclei is acceptable in present study.

\begin{figure}[htb]
\centering
\includegraphics[width=6cm]{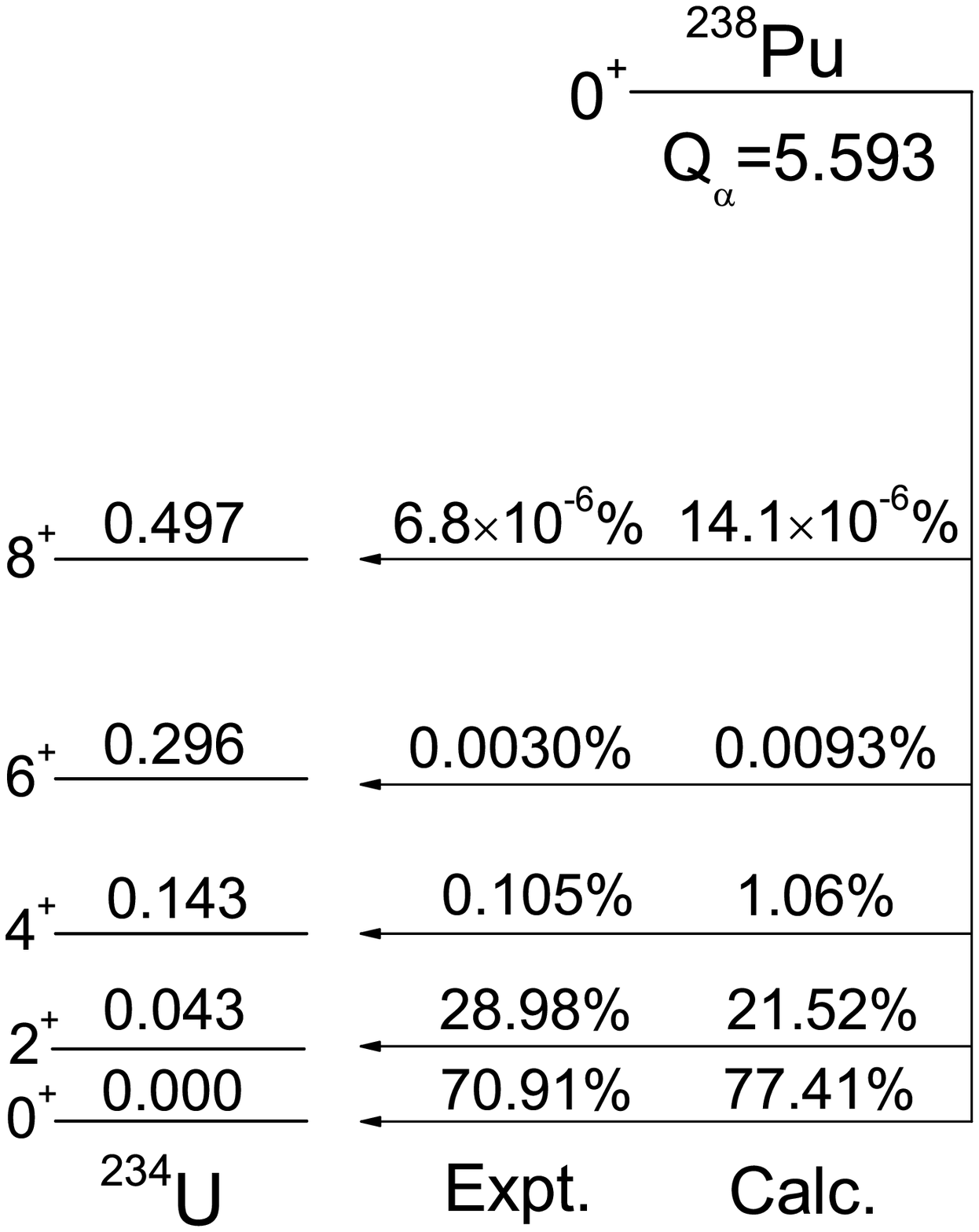}
\caption{The $\alpha$-decay to the rotational band of the ground-state of $^{238}$Pu.}
\end{figure}

\begin{figure}[htb]
\centering
\includegraphics[width=6cm]{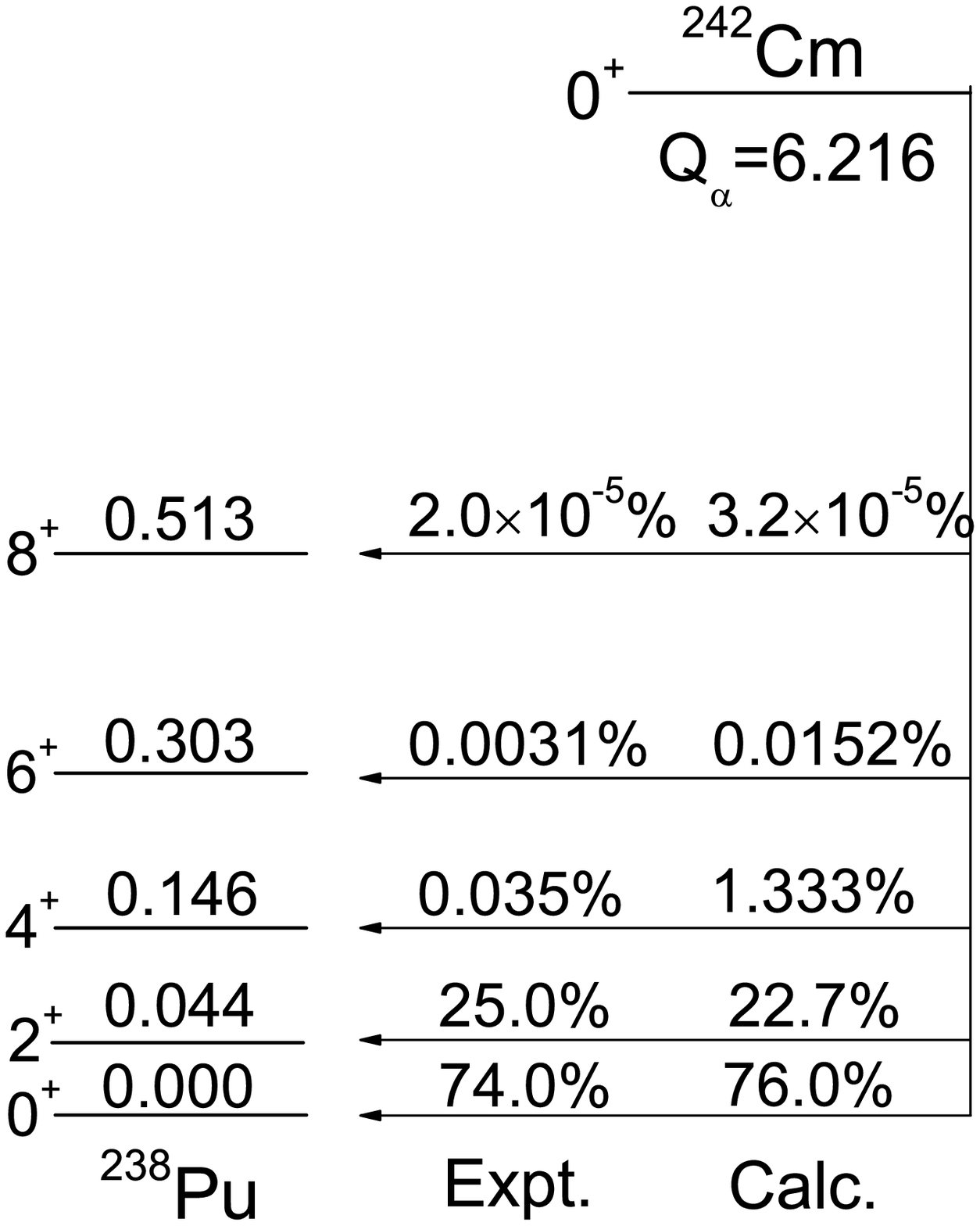}
\caption{The $\alpha$-decay to the rotational band of the ground-state of $^{242}$Cm.}
\end{figure}

\begin{figure}[htb]
\centering
\includegraphics[width=6cm]{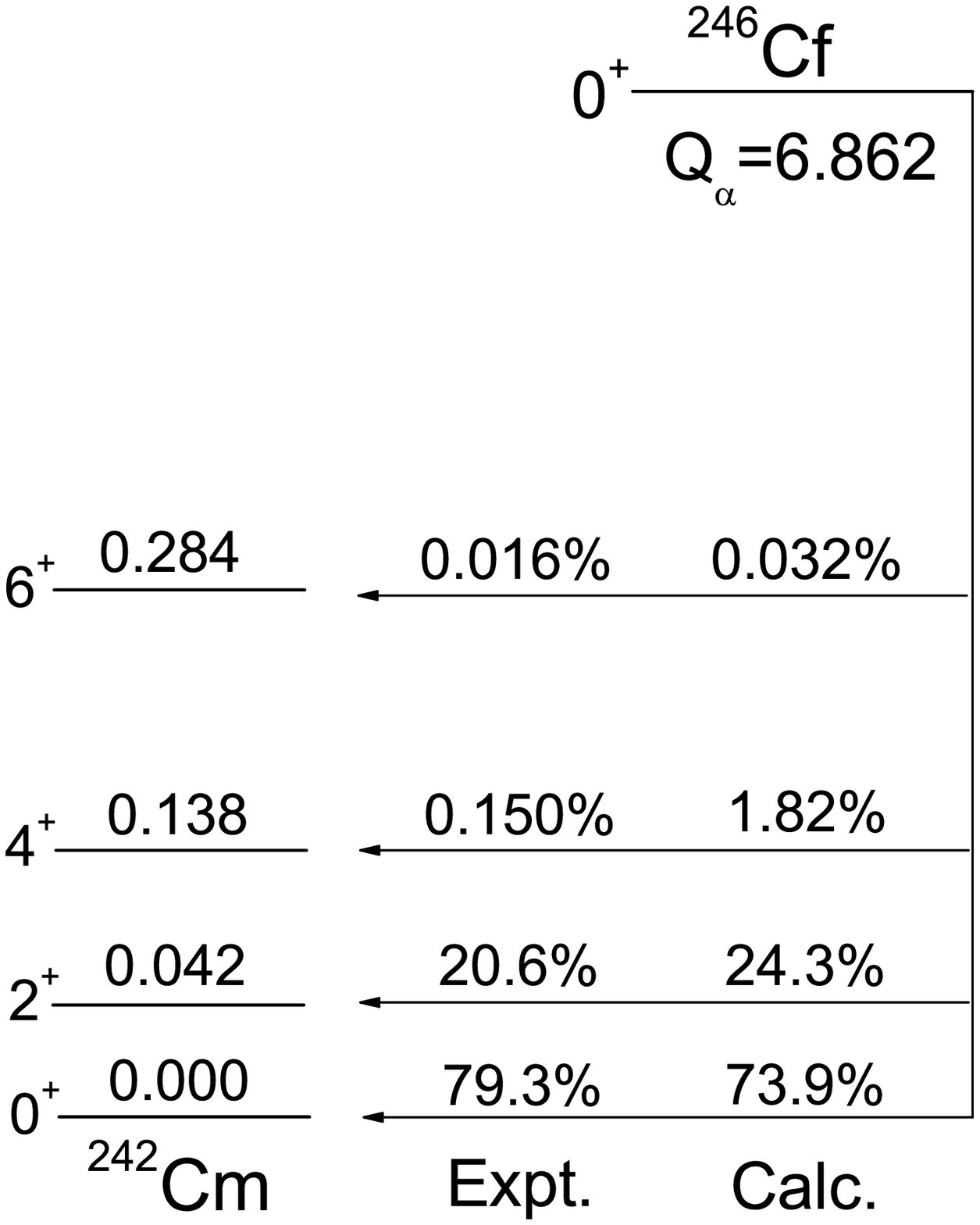}
\caption{The $\alpha$-decay to the rotational band of the ground-state of $^{246}$Cf.}
\end{figure}

We also perform a systematic calculation on the unfavored
$\alpha$-decays to the excited $0^+$ states of even-even
$\alpha$-emitters in the actinide region. Table I gives the
experimental and calculated branching ratios of
$\alpha$-transition to the excited 0$^{+}$ states for even-mass
isotopes of Th, U, Pu, and Cm. The experimental ground-state
branching ratio ($b_{g.s.}^{0 ^{+}}\%$) is used and its variation
ranges mainly from 67.4\% to 77.9\% for different nuclei in this
region \cite{fir}. However, the variation of the experimental
branching ratio to the excited $0^+$ state is relatively much
larger and its amplitude is as high as
0.0006\%/(5.1$\times$$10^{-7}$\%)$\approx$$10^3$ times (see Table
I).
\begin{table}[htb]
\centering \caption{Experimental and calculated branching ratios
of $\alpha$-decay to the excited 0$^{+}$ states of the daughter
nucleus. $\textrm{Q}_{\alpha}$ is the ground-state $\alpha$-decay
energy and $b_{g.s.}^{0 ^{+}}\%$ is the branching ratio of
$\alpha$-decay to the ground state of the daughter nucleus.
$\textrm{E}_{0+}^{*}$ is the excitation energy of the excited
$0^{+}$ state. $b_{e.s.}^{0 ^{+}}\%$ is the corresponding
experimental or theoretical $\alpha$-decay branching ratio.}
\begin{tabular}{ccccccc}
\hline \hline
         $\textrm{Nuclei}$
        &$\textrm{Q}_{\alpha}{(\textrm{MeV})}$
        &$b_{g.s.}^{0 ^{+}}\%{(\textrm{Expt.})}$
        &$\textrm{E}_{0+}^{*}{(\textrm{MeV})}$
        &$b_{e.s.}^{0 ^{+}}\%{(\textrm{Expt.})}$
        &$b_{e.s.}^{0 ^{+}}\%{(\textrm{Calc.})}$
        \\
\hline
$^{226}$Th      & 6.452 & 75.5\%  & 0.914 & 0.00034\%             & 0.00040\% \\
$^{228}$Th      & 5.520 & 71.1\%  & 0.916 & 1.8$\times$$10^{-5}$\% & 1.6$\times$$10^{-5}$\%\\
$^{230}$Th      & 4.770 & 76.3\%  & 0.825 & 3.4$\times$$10^{-6}$\% & 3.2$\times$$10^{-6}$\%\\
$^{232}$Th      & 4.083 & 77.9\%  & 0.721 & \#                     & 7.1$\times$$10^{-7}$\%\\
$^{230}$U       & 5.993 & 67.4\%  & 0.805 & 0.00030\%              & 0.00042\%\\
$^{232}$U       & 5.414 & 68.0\%  & 0.832 & 2.2$\times$$10^{-5}$\% & 3.6$\times$$10^{-5}$\%\\
$^{234}$U       & 4.859 & 71.4\%  & 0.635 & 2.6$\times$$10^{-5}$\% & 0.00025\%*\\
$^{236}$U       & 4.572 & 73.8\%  & 0.730 & \#                     & 7.5$\times$$10^{-6}$\%\\
$^{236}$Pu      & 5.867 & 69.3\%  & 0.691 & 0.0006\%               & 0.0016\% \\
$^{238}$Pu$^{1}$& 5.593 & 70.9\%  & 0.810 & 5$\times$$10^{-5}$\%   & 8$\times$$10^{-5}$\%\\
$^{238}$Pu$^{2}$& 5.593 & 70.9\%  & 1.045 & 1.2$\times$$10^{-6}$\% & 0.9$\times$$10^{-6}$\%\\
$^{240}$Pu      & 5.256 & 72.8\%  & 0.919 & 6.3$\times$$10^{-7}$\% & 2.3$\times$$10^{-6}$\%\\
$^{242}$Pu      & 4.983 & 77.5\%  & 0.926 & \#                     & 4.8$\times$$10^{-7}$\%\\
$^{242}$Cm$^{1}$& 6.216 & 74.0\%  & 0.942 & 5.2$\times$$10^{-5}$\% & 5.6$\times$$10^{-5}$\%\\
$^{242}$Cm$^{2}$& 6.216 & 74.0\%  & 1.229 & 5.1$\times$$10^{-7}$\% & 3.8$\times$$10^{-7}$\%\\
$^{244}$Cm$^{1}$& 5.902 & 76.4\%  & 0.861 & 1.55$\times$$10^{-4}$\%& 0.81$\times$$10^{-4}$\%\\
$^{244}$Cm$^{2}$& 5.902 & 76.4\%  & 1.089 & \#                     & 1.2$\times$$10^{-6}$\%\\
\hline \hline
\end{tabular}
\footnotetext{\# represents the cases where the experimental
branching ratio is still unknown.} \footnotetext{ * denotes the
cases where the calculated branching ratio deviates from the
experimental one.}
\end{table}
Therefore it is a challenging task to obtain a quantitative
agreement between experiment and theory. Unexpectedly, our simple
barrier penetration model with only one free parameter yields
results in good agreement with the experimental data. From the
last two columns of Table I, we can see that the calculated
branching ratios are very close to the experimental ones. The
experimental data are generally reproduced within a factor of 2
except for the case of $^{234}$U. The calculated value of
$^{234}$U is ten times larger than the experimental data and we
denote this abnormally large value by symbol * in Table I. We
notice that the excitation energy $\textrm{E}_{0+}^{*}$ in the
decay of $^{234}$U is significantly lower than those of
neighboring nuclei. Besides the first excited $0^+$ state, the
$\alpha$-transitions to the second excited $0^+$ state have also
been observed in experiment for some nuclei, such as $^{238}$Pu
and $^{242}$Cm. Our calculated branching ratios also agree with
the experimental ones in these cases. In Table I, the experimental
branching ratios to the first excited $0^+$ state have not been
measured yet for nuclei $^{232}$Th, $^{236}$U, and $^{242}$Pu
\cite{fir}. We list the corresponding predicted values for these
nuclei in Table I. Meanwhile, the predicted branching ratio to the
second excited $0^+$ state in decay of $^{244}$Cm is also given in
Table I. It is very interesting to compare these theoretical
predictions with future experimental observations.

\begin{table}[htb]
\centering \caption{The same as Table I, but for even isotopes of Rn,
Po, Pb and Hg.
The experimental data of $^{180}$Hg$-$$^{202}$Rn are taken from Ref.\cite{wau}.
The experimental data of $^{190}$Po are taken from Ref.\cite{and}.}
\begin{tabular}{ccccccc}
\hline \hline
         $\textrm{Nuclei}$
        &$\textrm{Q}_{\alpha}{(\textrm{MeV})}$
        &$b_{g.s.}^{0 ^{+}}\%{(\textrm{Expt.})}$
        &$\textrm{E}_{0+}^{*}{(\textrm{MeV})}$
        &$b_{e.s.}^{0 ^{+}}\%{(\textrm{Expt.})}$
        &$b_{e.s.}^{0 ^{+}}\%{(\textrm{Calc.})}$
        \\
\hline
$^{202}$Rn      & 6.775 & (80$-$100)\% & 0.816 & (1.4$-$1.8)$\times$$10^{-3}$\%  & (5.4$-$6.7)$\times$$10^{-3}$\%\\
$^{198}$Po      & 6.307 &  57\%        & 0.931 & 7.6$\times$$10^{-4}$\%          & 3.4$\times$$10^{-4}$\%\\
$^{196}$Po      & 6.657 &  94\%        & 0.769 & 2.1$\times$$10^{-2}$\%          & 1.1$\times$$10^{-2}$\%\\
$^{194}$Po      & 6.986 &  93\%        & 0.658 & 0.22\%                          & 7.3$\times$$10^{-2}$\%\\
$^{188}$Pb      & 6.110 & (3$-$10)\%   & 0.375 & (2.9$-$9.5)$\times$$10^{-2}$\%  & (2.9$-$9.7)$\times$$10^{-2}$\% \\
$^{186}$Pb      & 6.474 &   $<$100\%   & 0.328 & $<$0.20\%                       & 2.4\%*\\
$^{184}$Hg      & 5.658 &  1.25\%      & 0.478 & 2.0$\times$$10^{-3}$\%          & 1.9$\times$$10^{-3}$\%\\
$^{182}$Hg      & 5.997 &  8.6\%       & 0.422 & 2.9$\times$$10^{-2}$\%          & 4.4$\times$$10^{-2}$\%\\
$^{180}$Hg      & 6.257 &  33\%        & 0.443 & 2.6$\times$$10^{-2}$\%          & 0.18\%*\\
\hline
$^{190}$Po$^{1}$& 7.695 &  96.4\%      & 0.532 & 3.3\%                           & 6.4$\times$$10^{-1}$\%\\
$^{190}$Po$^{2}$& 7.695 &  96.4\%      & 0.650 & 0.3\%                           & 0.2\%\\
\hline
\hline
\end{tabular}
\footnotetext{ * denotes the cases where the calculated branching
ratio deviates from the experimental one.}
\end{table}

In Table II, we list the experimental and calculated branching
ratios to the excited $0^+$ states for even-mass isotopes of Rn,
Po, Pb and Hg. The hindered transitions ($\triangle$$\ell$=0) of
these nuclei involve complex particle-hole excitations above or
below the closed shell Z=82 \cite{woo}. Although the situation
becomes more complicated, it is seen from Table II that the
experimental results are reasonably reproduced by the simple
barrier penetration model. The $\alpha$-decay energies, the
ground-state branching ratios, and the excitation energies of the
nuclei in Table II are taken from the experimental values given by
Wauters \textit{et al.} \cite{wau} and by Andreyev \textit{et al.}
\cite{and}. It is found that the calculated branching ratios of
$^{202}$Rn, $^{190, 194-198}$Po, $^{188}$Pb and $^{182,184}$Hg are
consistent with the experimental data. For $^{186}$Pb and
$^{180}$Hg, the calculated values slightly deviate from the
experimental ones. The agreement may be further improved by taking
into account nuclear deformations and shape changes in these
nuclei \cite{woo}.

\section{Summary}

To conclude, we apply a simple barrier penetration approach to
calculate $\alpha$-decay branching ratios of even-even nuclei with
mass numbers 180$<$A$<$202 and A$\geq$224. We improve the original
$\alpha$-decay formula by taking into account the excitation
probability of the residual daughter nucleus. The calculated
branching ratios to the rotational band of the ground state of
even-even actinides are consistent with the experimental data. The
abnormal deviation of branching ratios to $4^+$ state is a common
phenomenon in present calculation and deserves further analysis.
The calculated branching ratios to the first and second excited
$0^+$ states of the daughter nucleus are also in agreement with
the available experimental values. Some predicted branching ratios
are given for the cases where the experimental values are still
unknown. It is hoped that our present barrier penetration approach
will serve as a good starting point for the microscopic study of
$\alpha$-decay phenomenon.

\

\begin{large}
\textbf{Acknowledgements}
\end{large}

This work is supported by National Natural Science Foundation of
China (No.10125521, No.10535010) and by 973 National Major State
Basic Research and Development of China (No.G2000077400).


\begin{thebibliography}{99}

\bibitem{gam}G. Gamov, Z. Phys. 51 (1928) 204.

\bibitem{con}E. U. Condon and R. W. Gurney, Nature 122 (1928) 439.

\bibitem{vio}V. E. Viola and G. T. Seaborg, J. Inorg. Nucl. Chem. 28 (1966) 741.

\bibitem{poe}D. N. Poenaru and M. Ivascu, J. Phys. 44 (1983) 791.

\bibitem{hat}Y. Hatsukawa, H. Nakahara and D. C. Hoffman, Phys. Rev. C 42 (1990) 674.

\bibitem{bro}B. A. Brown, Phys. Rev. C 46 (1992) 811.

\bibitem{buc}B. Buck, A. C. Merchant, and S. M. Perez,
Atomic Data and Nuclear Data Tables 54 (1993) 53.

\bibitem{roy}G. Royer, J. Phys. G: Nucl. Part. Phys. 26 (2000) 1149.

\bibitem{den}V. Yu. Denisov and H. Ikezoe, Phys. Rev. C 72 (2005) 064613.

\bibitem{moh}P. Mohr, Phys. Rev. C 73 (2006) 031301(R).

\bibitem{xu1}Chang Xu and Zhongzhou Ren, Phys. Rev. C 73 (2006) 041301(R).

\bibitem{xu2}Chang Xu and Zhongzhou Ren, Phys. Rev. C 74 (2006) 014304.

\bibitem{fir}R. B. Firestone, V. S. Shirley, C. M. Baglin, S. Y. Frank Chu and
J. Zipkin, Table of Isotopes, 8th ed. Wiley Interscience, New
York, 1996.

\bibitem{mol}P. M\"oller, J. R. Nix and K. -L. Kratz,
Atomic Data and Nuclear Data Tables 66 (1997) 131.

\bibitem{wau}J. Wauters, N. Bijnens, P. Dendooven, M. Huyse,
H. Y. Hwang, G. Reusen, J. von Schwarzenberg,
P. Van Duppen, R. Kirchner and E. Roeckl, Phys. Rev. Lett. 72 (1994) 1329.

\bibitem{and}A. N. Andreyev, M. Huyse, P. Van Duppen, L. Weissman,
D. Ackermann, J. Gerl, F. P. He$\ss$erger, S. Hofmann, A.
Kleinb\"{o}hl, G. M\"{u}nzenberg, S. Reshitko, C. Schlegel, H.
Schaffner, P. Cagarda, M. Matos, S. Saro, A. Keenan, C. Moore, C.
D. O'Leary, R. D. Page, M. Taylor, H. Kettunen, M. Leino, A.
Lavrentiev, R. Wyss and K. HeydeI, Nature 405 (2000) 430.

\bibitem{ric}J. D. Richards, C. R. Bingham, Y. A. Akovali, J. A. Becker, E. A. Henry, P. Joshi,
J. Kormicki, P. F. Mantica, K. S. Toth, J. Wauters and E. F. Zganjar,
Phys. Rev. C 56 (1996) 2041.

\bibitem{del}D. S. Delion, A. Florescu, M. Huyse, J. Wauters,
P. Van Duppen(ISOLDE Collaboration), A. Insolia and R. J. Liotta,
Phys. Rev. Lett. 74 (1995) 3939.

\bibitem{kar}D. Karlgren, R. J. Liotta, R. Wyss, M. Huyse,
K. Van de Vel and P. Van Duppen,
Phys. Rev. C 73 (2006) 064304.

\bibitem{asa}M. Asai, K. Tsukada, S. Ichikawa, M. Sakama, H. Haba, I. Nishinaka, Y. Nagame,
S. Goto, Y. Kojima, Y. Oura and M. Shibata,
Phys. Rev. C 73 (2006) 067301.

\bibitem{xu}Chang Xu and Zhongzhou Ren, Nucl. Phys. A753 (2005) 174;
A760 (2005) 303.

\bibitem{gam2}G. Gamov and C. L. Critchfueld, Theory of atomic nucleus and
nuclear energy-sources, Vol. III, The Clarendon Press, Oxford, 1949.

\bibitem{ras} J. O. Rasmussen, in Alpha-, Beta-, and Gamma-Ray Spectroscopy,
Vol. I, North Holland, Amsterdam, 1965.

\bibitem{boh}A. Bohr and B. R. Mottelson, Nuclear Structure, Vol.
II, World Scientific, Singapore, 1998.

\bibitem{woo}J. L. Wood, K. Heyde, W. Nazarewicz, M. Huyse and P. van Duppen,
Phys. Rep. 215 (1992) 101.

\end{thebibliography}
\end{document}